%% file: main.tex
\documentclass{Interspeech}

\interspeechcameraready 

\title{SoundSculpt: Direction and Semantics Driven Ambisonic Target Sound Extraction}

\author[affiliation={1}]{Tuochao}{Chen$^{*}$}
\author[affiliation={2}]{D}{Shin$^{**}$}
\author[affiliation={3}]{Hakan}{Erdogan}
\author[affiliation={2}]{Sinan}{Hersek}

\affiliation{}{University of Washington}{}
\affiliation{}{Google AR}{}
\affiliation{}{Google DeepMind}{}
\email{tuochao@cs.washington.edu, deshin@google.com, hakanerdogan@google.com, shersek@google.com}

\def\thefootnote{\fnsymbol{footnote}}
\usepackage{comment}
\usepackage{gensymb}
\usepackage{multirow}
\usepackage{siunitx}
\usepackage{textcomp}
\usepackage{mathtools}
\usepackage[symbol]{footmisc}

\DeclarePairedDelimiter\abs{\lvert}{\rvert}

\newcommand{\smallsim}{\smallsym{\mathrel}{\sim}}

\makeatletter
\newcommand{\smallsym}[2]{#1{\mathpalette\make@small@sym{#2}}}
\newcommand{\make@small@sym}[2]{%
  \vcenter{\hbox{$\m@th\downgrade@style#1#2$}}%
}
\newcommand{\downgrade@style}[1]{%
  \ifx#1\displaystyle\scriptstyle\else
    \ifx#1\textstyle\scriptstyle\else
      \scriptscriptstyle
  \fi\fi
}
\makeatother

\begin{document}

\maketitle

\renewcommand{\thefootnote}{\fnsymbol{footnote}}
\footnote[0]{* Work done during internship in Google AR.}
\footnote[0]{** Now at Google DeepMind.}

\begin{abstract}
\input{abstract}    
\end{abstract}

\section{Introduction}
\input{introduction}

\section{Method}

\subsection{Problem Setup}
\input{problem_setup}

\subsection{Model Architecture}
\input{model_architecture}

\subsection{Dataset Preparation}
\input{dataset_preparation}

\subsection{Training Setup}
\input{training_setup}

\subsection{Baseline Methods}
\input{baseline_methods}

\section{Results and Discussion}
\input{results_and_discussion}

\section{Conclusion}
\input{conclusion}

\bibliographystyle{IEEEtran}
\bibliography{mybib}

\end{document}

%% file: abstract.tex
This paper introduces SoundSculpt, a neural network designed to extract target sound fields from ambisonic recordings. SoundSculpt employs an ambisonic-in-ambisonic-out architecture and is conditioned on both spatial information (e.g., target direction obtained by pointing at an immersive video) and semantic embeddings (e.g., derived from image segmentation and captioning). Trained and evaluated on synthetic and real ambisonic mixtures, SoundSculpt demonstrates superior performance compared to various signal processing baselines. Our results further reveal that while spatial conditioning alone can be effective, the combination of spatial and semantic information is beneficial in scenarios where there are secondary sound sources spatially close to the target. Additionally, we compare two different semantic embeddings derived from a text description of the target sound using text encoders.

%% file: introduction.tex
In recent years, spatial audio has become increasingly popular due to the introduction of new VR/AR headsets. Ambisonics, a popular spatial audio format, enables binaural rendering of head-tracked spatial audio and is commonly used in VR \cite{zotter2019ambisonics}. Ambisonic signals can be recorded using spherical microphone arrays \cite{rafaely2015fundamentals} or emerging techniques that employ arbitrary microphone arrays, such as those found in AR glasses \cite{gayer2024ambisonics}.

\begin{figure}[t]
  \centering
  \includegraphics[width=\linewidth]{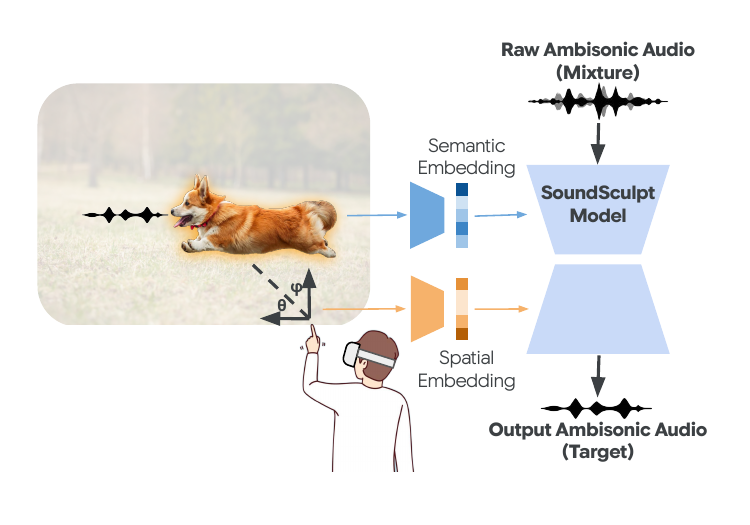}
  \caption{Overview of SoundSculpt, a spatially and semantically conditioned target ambisonic signal extraction model.}
  \label{fig:overview}
\end{figure}

In this paper, we present SoundSculpt, a method to extract a target ambisonic sound field from an ambisonic mixture recording. We envision a scenario where an individual watches a $180^{o}$ or $360^{o}$ panoramic video with accompanying head tracked spatial audio on a VR headset (Figure \ref{fig:overview}). For example, the video might feature a street musician playing the guitar and singing amidst a busy cityscape, with various other ambient sounds like people talking and traffic. The user points to the guitar, and all other sounds in the scene vanish, leaving only the guitar sound, preserving its spatial and reverb characteristics.

The main component of SoundSculpt is an ambisonic-in-ambisonic-out neural network conditioned by a target direction and a semantic embedding that describes the sound to be isolated. The target direction can be provided by the user pointing to an immersive video. A semantic representation of the target sound can be obtained using regional image segmentation at the target direction, followed by the computation of a semantic embedding of the segmented region.

Signal processing based techniques exist to estimate a target ambisonic signal. In \cite{kronlachner2014spatial} an ambisonic domain transformation is proposed for directional loudness modification. The loudness of sounds outside of a spherical cap centered at the target direction can be suppressed using this method. Furthermore, \cite{herzog2019direction} and \cite{herzog2020direction} investigate beamforming and matrix multi-channel Wiener filtering based methods to estimate a target ambisonic signal. However, if there is another sound source close to the desired target, these techniques might not perform effectively. 

This raises the question of whether information other than spatial direction can be employed as a conditioning factor for extracting target ambisonic signals.

Semantically conditioned neural networks have shown promising results to extract target signals from mono or binaural recordings. Kilgour et al. \cite{kilgour2022text} present a semantically conditioned audio-to-audio UNet architecture designed to isolate a desired audio source from mono waveforms. The semantic condition can be derived by feeding the text description of the target sound to a text encoder or a short audio snippet of the target source to an audio encoder. \cite{veluri2023real} describes a computationally efficient neural network for extracting target sounds from mono waveforms. Furthermore, Veluri et al. \cite{veluri2023semantic} explore the preservation of binaural cues during target sound extraction from binaural audio. Both of the latter approaches utilize the class label of the target sound for model conditioning.

Neural networks have also shown promising results in region-wise sound extraction. For instance, Yang et al. proposed a model that separates target speech from interfering sources using two microphones, focusing on different angular regions \cite{yang2024binaural}. Gu et al. introduced a method for extracting active target sounds within a user-defined spatial region, utilizing multi-channel audio input and region features to produce an estimate of the target signal \cite{gu2024rezero}.

Drawing inspiration from these methods, we introduce a first order ambisonic-to-ambisonic UNet architecture designed to extract target ambisonic sounds from complex ambisonic mixtures, leveraging both spatial and semantic conditioning. To evaluate SoundSculpt's performance, we curate a comprehensive dataset of synthetic and real ambisonic recordings and conduct extensive experiments comparing different conditioning strategies (spatial, semantic, and combined). Our results demonstrate that SoundSculpt significantly outperforms traditional signal processing baselines in terms of scale-invariant signal-to-distortion ratio improvement (SI-SDRi).  Furthermore, we analyze the benefits of incorporating both spatial and semantic cues, revealing their synergistic contribution to target sound extraction. Finally, we investigate the impact of different semantic embedding techniques, comparing the contrastively trained SoundWords \cite{kilgour2022text} encoder with a widely-used BERT model \cite{kenton2019bert}, and discuss their relative effectiveness.

%% file: problem_setup.tex
Let $a(t, \theta, \phi)$ be the plane wave amplitude distribution describing an incident sound field. We denote the azimuth angle as $\theta \in [0, 2\pi]$ and the elevation angle as $\phi \in [-\frac{\pi}{2},\frac{\pi}{2}]$. The ambisonic signal, $\boldsymbol{x}(t) = [x_{00}(t),...,x_{nm}(t),...x_{NN}(t)]^{T} \in \mathbb{R}^{(N+1)^{2}}$, represents this sound field using a spherical harmonics (SH) expansion up to SH (or ambisonic) order $N$:
\begin{align}
  a(t, \theta, \phi) \approx \sum_{n=0}^{N} \sum_{m=-n}^{n} {x}_{nm}(t)Y_{n}^{m}(\theta, \phi)
  \label{equation:sh_expansion}
\end{align}
where $Y_{n}^{m}$ are real valued SH basis functions, $n$ and $m$ are the SH order and degree respectively \cite{kronlachner2014spatial}. In this work we use the SN3D SH basis described in \cite{gorzel2019efficient}. Additionally, we use first order ambisonics (FOA) where $N=1$.

We assume that an ambisonic recording $\boldsymbol{x}(t)$ is the combination of a target sound field and other sounds in the scene:
\begin{align}
  \boldsymbol{x}(t) = \boldsymbol{y}(t) + \boldsymbol{n}(t)
  \label{equation:signal_model}
\end{align}
where $\boldsymbol{y}(t) \in \mathbb{R}^{(N+1)^{2}}$ is the ambisonic signal representing the target sound field. Similarly $\boldsymbol{n}(t)$ represents the sound field due to everything else in the scene.

Our goal is to estimate the target ambisonic signal $\boldsymbol{\hat{y}}(t)$ given the ambisonic recording $\boldsymbol{x}(t)$. To that end, we aim to design a network that takes $\boldsymbol{x}(t)$, the source direction and a semantic embedding vector that describes the target sound, as inputs. We propose extracting the semantic embedding - denoted using $\boldsymbol{v} \in \mathbb{R}^{K}$ - from a text description of the target sound using a text encoder. We denote the target sound's azimuth and elevation angles of arrival as $\theta_{s}$ and $\phi_{s}$. We aim to design an estimator to extract the target ambisonic signal:
\begin{align}
  \boldsymbol{\hat{y}} = f(\boldsymbol{x}, \theta_{s}, \phi_{s}, \boldsymbol{v})
  \label{equation:estimator}
\end{align}
where $T$ is the number of time samples, $\boldsymbol{x} \in \mathbb{R}^{(N+1)^{2} \times T}$ is a segment of the ambisonic recording and $\boldsymbol{\hat{y}} \in \mathbb{R}^{(N+1)^{2} \times T}$ is the corresponding estimated target. We assume that the target source doesn't move during the duration of the segment.

We hypothesize that in scenarios where the target sound originates from a direction (e.g. a plane wave) with minimal interference from nearby directions, knowing the source direction alone should suffice for effective target ambisonic signal estimation. However, the presence of interfering sounds from nearby directions may necessitate additional information beyond the direction. Given the promising results of semantic conditioning for target sound extraction in mono and binaural signals, we anticipate that incorporating a semantic input will be beneficial in such scenarios \cite{kilgour2022text, veluri2023real, veluri2023semantic}. 

To enhance flexibility and avoid limiting SoundSculpt to predefined sound classes, we utilize a semantic embedding based on a text description rather than discrete class labels as employed in \cite{veluri2023semantic}. This approach allows for a richer and more nuanced representation of the target sound, enabling the model to generalize to a wide range of acoustic events.

%% file: model_architecture.tex
Our model draws inspiration from the semantically conditioned mono-to-mono U-Net architecture in \cite{kilgour2022text}, extending it to a first order ambisonic-to-ambisonic (4-channel) U-Net with semantic and/or spatial conditioning.

We derive the semantic embedding vector by feeding the target sound's text description into a text encoder model. One of the encoders we consider is SoundWords from \cite{kilgour2022text}, which is trained to generate a joint text-sound embedding through contrastive learning. We also investigate a widely-used BERT encoder trained in a self-supervised manner on the Wikipedia + BooksCorpus datasets \cite{google2023bert}\cite{kenton2019bert}.

To condition our model, we project the semantic embedding and target direction (or the spatial condition - consisting of the azimuth and elevation angles) separately to vectors of size 128 that are then added together. This is then incorporated into the model using feature-wise linear modulation (FiLM) \cite{perez2018film}, applied to the layers within the encoder and decoder. Furthermore, we experiment with using only the semantic condition and only the spatial conditions. Similar to \cite{kilgour2022text}, we employ a symmetric encoder-decoder U-net network with skip-connections. The encoder employs a convolutional architecture with an initial depth of 32, progressively increasing by multiples of 8, 8, 2, and 2 through 7-layer ResNet blocks \cite{he2016deep}. The decoder mirrors this structure.

%% file: dataset_preparation.tex
\begin{figure*}[ht]
  \centering
  \includegraphics[width=\textwidth]{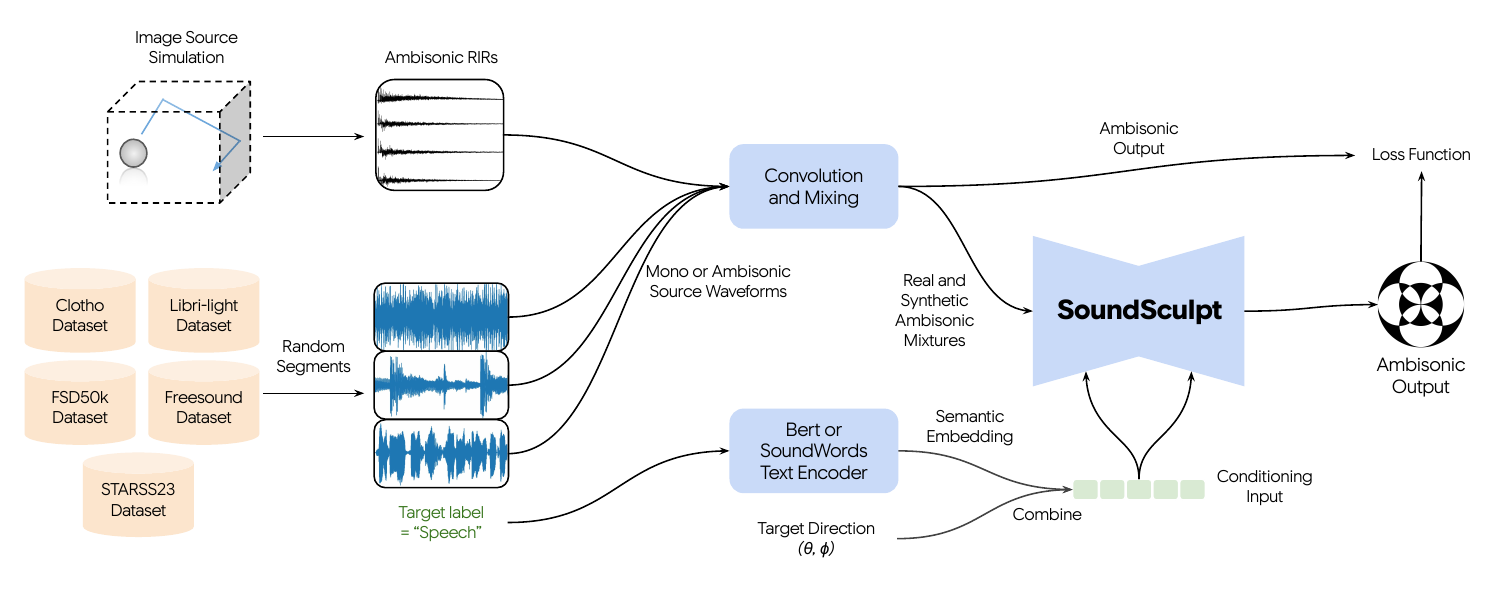}
  \caption{Overview of dataset preparation and model training.}
  \label{fig:dataset_prep}
\end{figure*}

We train and test our models on both synthetic and real ambisonic data as shown in Figure \ref{fig:dataset_prep}. To create synthetic data we first use the image source method to generate 400k ambisonic room impulse responses (RIRs) for training and 80k for evaluation \cite{allen1979image}. To generate each ambisonic RIR, we first generate a random shoebox shaped room where the room dimensions can randomly vary between 2-15 meters. We place an ambisonic receiver near the center of the room with some position randomization. We randomly position four sound sources within the room, where the first source is the target. Each source's direction relative to the receiver is sampled randomly from a uniform distribution on the unit sphere \cite{muller1959note}. The distance between each source and the receiver is randomly sampled from the range 0.6-5 meters, ensuring that all sources fall inside the room.

Random material types are assigned to the floor, ceiling and walls of the room and FIR filters representing the acoustic characteristics of these materials are used during simulation. Jitter is introduced to the positions of the image sources to mitigate sweeping echoes in the RIRs \cite{de2015modeling}. 

We conduct the simulations in the frequency domain. For a given sound source, denoting the number of image source paths to the receiver as $P$ we compute the ambisonic room transfer function for SH order $n$ and degree $m$ at frequency $f$ as:
\begin{align}
  r_{nm}(f) = \sum_{p=1}^{P} I_{p}(f) e^{-j 2 \pi f \tau_{p}}
  Y^{m}_{n}(\theta_{p}, \phi_{p})
\end{align}
where, for each path $p$, $I_{p}(f) \in \mathbb{C}$ is a multiplier related to distance attenuation and wall filtering, $\tau_{p}$ is the travel delay and $(\theta_{p}, \phi_{p})$ is the direction of arrival. This is equivalent to encoding each image source path to ambisonics \cite{gorzel2019efficient}. Simulation is distributed amongst 100 NVIDIA Tesla V100 GPUs for parallel computing.

A synthetic training dataset of 1.2 million input/target pairs was created by convolving and mixing mono audio sources with randomly selected ambisonic RIRs. Source waveforms were randomly selected from Libri-Light \cite{kahn2020libri}, Clotho \cite{drossos2020clotho}, FSD50k \cite{fonseca2021fsd50k}, and a collection of $\sim$78k CC0-licensed Freesound clips \cite{freesound} that are not part of FSD50k. Random source gains and silencing of secondary sources were applied to increase diversity.

A similar process was used to create a synthetic test dataset of 120k pairs, with speech segments from LibriTTS dev-clean split \cite{zen2019libritts} and arbitrary sounds from Clotho and FSD50k's test splits.

Target source descriptions were derived from dataset labels (FSD50k), captions (Clotho), or assigned based on the source dataset (e.g. variations of the phrase "human speech" for Libri-Light and LibriTTS). The additional Freesound clips were labeled using keyword matching against the AudioSet ontology \cite{gemmeke2017audio} and a sound event classifier. These descriptions were used to extract semantic embeddings for the target sounds.

An additional 400k training input/target pairs were created by sampling and mixing segments from the development set of the STARSS23 dataset \cite{shimada2024starss23}. This dataset consists of real FOA recordings obtained using a high-resolution 32-channel spherical microphone array (Eigenmike EM32). The sound event and directions are annotated. We first sample 4.096s clips consisting of single sound events where the direction is static. We randomly select two segments, one for the target source and one for a secondary source. We apply random gains to these segments and mix them. We randomize the source directions for each segment, by uniformly sampling a random direction on the unit sphere and rotating the FOA signal to match this direction of arrival via multiplications with rotation matrices \cite{zotter2019ambisonics}. The target segment's sound event label serves as its description. We similarly create 40k pairs to form a test split of STARSS23 mixtures, using the evaluation set of the STARSS23 dataset. During mixture creation, we excluded the sound class "domestic sounds" from consideration as a target source due to its broad and encompassing nature. 

All input/target pairs we formed have a sample rate of 16kHz and duration of 4.096s. Additionally, for both datasets, with 50\% probability we placed a secondary source that has a different semantic description than the target, near the target source (within $\pm$15° azimuth and elevation) to generate challenging cases.

\begin{table*}[!th]
\centering
\caption{Comparison of different target ambisonic signal extraction algorithms. We present SI-SDRi (averaged across channels) evaluated and averaged across the test split of the STARSS23 mixtures and synthetic datasets. Results are shown for all segments, segments where there is at least one secondary source within \ang{15}  of the target direction (only close secondary) and segments where there are no secondary sources within \ang{15} (no close secondary).}
\label{table:results}
\begin{tabular}{|l|c|c|c|c|}
\hline
\multicolumn{1}{|c|}{\textbf{\shortstack{Dataset\\Name}}} & \multicolumn{1}{c|}{\textbf{\shortstack{Algorithm Name}}} & \multicolumn{1}{c|}{\textbf{\shortstack{SI-SDRi [dB]\\(all)}}} & \multicolumn{1}{c|}{\textbf{\shortstack{SI-SDRi [dB]\\(only close secondary)}}} & \multicolumn{1}{c|}{\textbf{\shortstack{SI-SDRi [dB]\\(no close secondary)}}} \\

\hline
\multirow{10}{*}[2.3ex]{\shortstack{STARSS23\\Mixtures}} & Ambisonic Loudness Modification & -6.71 & -7.84 & -6.28 \\
 & Max-DI Beamform-and-project & -6.36 & -7.95 & -5.75 \\
 & Max-$\bm{r}_{E}$ Beamform-and-project & -6.97 & -8.48 & -6.40 \\
 & Bert Condition & -1.80 & -0.14 & -2.44 \\
 & Soundwords Condition & -2.13 & -0.59 & -2.72 \\
 & Spatial Condition & 2.44 & -0.59 & \textbf{3.59} \\
 & Spatial and Bert Condition & 0.73 & -0.20 & 1.09 \\
 & Spatial and Soundwords Condition & \textbf{2.47} & \textbf{1.19} & 2.96 \\

\hline
\multirow{10}{*}[2.2ex]{ Synthetic} & Ambisonic Loudness Modification & -1.43 & -2.32 & -1.16 \\
 & Max-DI Beamform-and-project & -1.56 & -2.83 & -1.17 \\
 & Max-$\bm{r}_{E}$ Beamform-and-project & -1.59 & -2.87 & -1.19 \\
 & Bert Condition & 4.03 & 4.18 & 3.99 \\
 & Soundwords Condition & 3.62 & 3.81 & 3.56 \\
 & Spatial Condition & 9.36 & 6.03 & 10.38  \\
 & Spatial and Bert Condition & 10.32 & 8.06 & 11.02 \\
 & Spatial and Soundwords Condition & \textbf{10.73} & \textbf{8.81} & \textbf{11.31} \\

\hline 
\end{tabular}
\end{table*}

%% file: training_setup.tex
Each SoundSculpt model was trained on 16 Google Cloud TPU v3 cores for at least 500k steps (batch size 128, learning rate $10^{-4}$ using the Adam optimizer). Each batch comprised 75\% of synthetic and 25\% of STARSS23 mixtures, this ratio was empirically determined. Training emphasized synthetic mixtures because the synthetic dataset offered more diversity and a larger volume of data. The loss function was the $\ell_{1}$ distance between the complex short-time Fourier transforms (STFT, window length 1024, hop length 256) of the target and output, averaged across the FOA channels. 

In more detail, we denote the STFT of the target signal as $\boldsymbol{Y} \in \mathbb{C}^{{L} \times {F} \times (N+1)^{2}}$ where $L$ is the number of time steps, $F=513$ is the number of real-FFT frequency bins and $N=1$ is the ambisonic order. Denoting the estimated signal's STFT as $\boldsymbol{\hat{Y}} \in \mathbb{C}^{{L} \times {F} \times (N+1)^{2}}$ we compute the loss as:
\begin{align}
  \mathcal{L} = \frac{1}{(N+1)^{2}}\sum_{c=1}^{(N+1)^{2}}\sum_{l=1}^{L} \,\,\, \sum_{f=1}^{F} \,\,\, \abs{\boldsymbol{Y}[l, f, c] - \boldsymbol{\hat{Y}}[l, f, c]}
  \label{equation:per_channel_loss}
\end{align}

%% file: baseline_methods.tex
SoundSculpt's performance is compared against three baseline signal processing algorithms. The first, directional loudness modification, employs an ambisonic transformation to preserve the loudness of the spatial signal within a spherical cap, while attenuating signals outside this region \cite{kronlachner2014spatial}. The center direction of the spherical cap is set to the target sound direction. For the cap spread, we empirically determined a value of 60\degree through experimentation on a small validation set.

Two additional baseline approaches, similar to the beamform-and-project method described by Herzog et al. \cite{herzog2020direction}, were implemented. This technique estimates source signals using beamformers and then projects these estimates back to the ambisonic domain. While the original method employs multi-speaker multichannel Wiener filters, requiring estimation of the power spectral density of both the source signals and the noise, we instead utilize the signal-independent max-DI and max-$\bm{r}_{E}$ beamformers \cite{lluis2023direction}. A single beamformer steered toward the target's direction is used.

%% file: results_and_discussion.tex
Results are presented in Table \ref{table:results}. We present SI-SDRi (averaged across channels) for the baseline algorithms and SoundSculpt models that use various types of conditioning. SI-SDRi is averaged across the respective datasets. We evaluate the algorithms on the test split of the STARSS23 mixtures and the synthetic dataset. In addition to presenting results for the whole dataset, we also provide results for segments where there's at least one secondary source within \ang{15} (great circle distance) of the target direction and for segments where there aren't any secondary sources within \ang{15} of the target direction.

For both the STARSS23 mixtures and the synthetic datasets, all neural network-based approaches significantly outperform the signal processing approaches, even when using only semantic conditioning.  On the synthetic data, SoundSculpt models using spatial + semantic conditions outperform both spatial-only and semantic-only conditioning, with little difference between BERT and SoundWords encoders. However, we note that SoundSculpt models have approximately 31.7 million parameters, whereas the signal processing methods are inherently more computationally efficient.

On the synthetic dataset, semantic-only conditioning exhibits a small SI-SDRi difference ($\smallsim$0.2dB) between segments with and without a nearby secondary source. Models using spatial-only or combined semantic and spatial conditioning perform worse when a secondary source is close to the target direction. In the segments with a nearby secondary source, the combined semantic and spatial models outperform the spatial-only SoundSculpt. While the combined models also outperform the spatial-only model when no secondary source is nearby, the performance gain is smaller.

The STARSS23 mixtures show lower SI-SDRi values overall compared to the synthetic datasets, possibly due to imprecision in direction and sound event labels or external noise in the extracted segments. Unlike the synthetic dataset, there is little difference between spatial-only and spatial + SoundWords conditioning, potentially due to fewer secondary sources in the STARSS23 mixtures.

When there's no secondary source close to the target, spatial-only conditioning slightly outperforms spatial + SoundWords by $\smallsim$0.6dB. However, with a close secondary source, spatial + SoundWords outperforms spatial-only by $\smallsim$1.8dB, demonstrating the benefits of combined conditioning on both realistic and synthetic ambisonic datasets. On the STARSS23 mixtures, spatial + BERT conditioning is outperformed by spatial + SoundWords, which may be because the BERT encoder is trained solely on text, whereas SoundWords is trained on both text and audio, likely resulting in a better alignment between the two modalities.

%% file: conclusion.tex
This paper introduced SoundSculpt, a neural network model for target sound extraction from ambisonic recordings, leveraging both spatial and semantic conditioning.  SoundSculpt consistently outperformed traditional signal processing baselines. The combination of spatial and semantic information proved beneficial in complex acoustic scenes with interfering sources near the target. Future work could explore alternative network architectures and different semantic embedding techniques.

%% file: main.bbl
\begin{thebibliography}{10}
\providecommand{\url}[1]{#1}
\csname url@samestyle\endcsname
\providecommand{\newblock}{\relax}
\providecommand{\bibinfo}[2]{#2}
\providecommand{\BIBentrySTDinterwordspacing}{\spaceskip=0pt\relax}
\providecommand{\BIBentryALTinterwordstretchfactor}{4}
\providecommand{\BIBentryALTinterwordspacing}{\spaceskip=\fontdimen2\font plus
\BIBentryALTinterwordstretchfactor\fontdimen3\font minus \fontdimen4\font\relax}
\providecommand{\BIBforeignlanguage}[2]{{%
\expandafter\ifx\csname l@#1\endcsname\relax
\typeout{** WARNING: IEEEtran.bst: No hyphenation pattern has been}%
\typeout{** loaded for the language `#1'. Using the pattern for}%
\typeout{** the default language instead.}%
\else
\language=\csname l@#1\endcsname
\fi
#2}}
\providecommand{\BIBdecl}{\relax}
\BIBdecl

\bibitem{zotter2019ambisonics}
F.~Zotter and M.~Frank, \emph{Ambisonics: A practical 3D audio theory for recording, studio production, sound reinforcement, and virtual reality}.\hskip 1em plus 0.5em minus 0.4em\relax Springer Nature, 2019.

\bibitem{rafaely2015fundamentals}
B.~Rafaely, \emph{Fundamentals of spherical array processing}.\hskip 1em plus 0.5em minus 0.4em\relax Springer, 2015, vol.~8.

\bibitem{gayer2024ambisonics}
Y.~Gayer, V.~Tourbabin, Z.~Ben-Hur, J.~Donley, and B.~Rafaely, ``Ambisonics encoding for arbitrary microphone arrays incorporating residual channels for binaural reproduction,'' \emph{arXiv preprint arXiv:2402.17362}, 2024.

\bibitem{kronlachner2014spatial}
M.~Kronlachner and F.~Zotter, ``Spatial transformations for the enhancement of ambisonic recordings,'' in \emph{Proceedings of the 2nd International Conference on Spatial Audio, Erlangen}, 2014.

\bibitem{herzog2019direction}
A.~Herzog and E.~A. Habets, ``Direction preserving wiener matrix filtering for ambisonic input-output systems,'' in \emph{ICASSP 2019-2019 IEEE International Conference on Acoustics, Speech and Signal Processing (ICASSP)}.\hskip 1em plus 0.5em minus 0.4em\relax IEEE, 2019, pp. 446--450.

\bibitem{herzog2020direction}
A.~Herzog and E.~A.~P. Habets, ``Direction and reverberation preserving noise reduction of ambisonics signals,'' \emph{IEEE/ACM Transactions on Audio, Speech, and Language Processing}, vol.~28, pp. 2461--2475, 2020.

\bibitem{kilgour2022text}
K.~Kilgour, B.~Gfeller, Q.~Huang, A.~Jansen, S.~Wisdom, and M.~Tagliasacchi, ``Text-driven separation of arbitrary sounds,'' \emph{arXiv preprint arXiv:2204.05738}, 2022.

\bibitem{veluri2023real}
B.~Veluri, J.~Chan, M.~Itani, T.~Chen, T.~Yoshioka, and S.~Gollakota, ``Real-time target sound extraction,'' in \emph{ICASSP 2023-2023 IEEE International Conference on Acoustics, Speech and Signal Processing (ICASSP)}.\hskip 1em plus 0.5em minus 0.4em\relax IEEE, 2023, pp. 1--5.

\bibitem{veluri2023semantic}
B.~Veluri, M.~Itani, J.~Chan, T.~Yoshioka, and S.~Gollakota, ``Semantic hearing: Programming acoustic scenes with binaural hearables,'' in \emph{Proceedings of the 36th Annual ACM Symposium on User Interface Software and Technology}, 2023, pp. 1--15.

\bibitem{yang2024binaural}
Y.~Yang, G.~Sung, S.-F. Shih, H.~Erdogan, C.~Lee, and M.~Grundmann, ``Binaural angular separation network,'' in \emph{ICASSP 2024-2024 IEEE International Conference on Acoustics, Speech and Signal Processing (ICASSP)}.\hskip 1em plus 0.5em minus 0.4em\relax IEEE, 2024, pp. 1201--1205.

\bibitem{gu2024rezero}
R.~Gu and Y.~Luo, ``Rezero: Region-customizable sound extraction,'' \emph{IEEE/ACM Transactions on Audio, Speech, and Language Processing}, 2024.

\bibitem{kenton2019bert}
J.~D. M.-W.~C. Kenton and L.~K. Toutanova, ``Bert: Pre-training of deep bidirectional transformers for language understanding,'' in \emph{Proceedings of naacL-HLT}, vol.~1.\hskip 1em plus 0.5em minus 0.4em\relax Minneapolis, Minnesota, 2019, p.~2.

\bibitem{gorzel2019efficient}
M.~Gorzel, A.~Allen, I.~Kelly, J.~Kammerl, A.~Gungormusler, H.~Yeh, and F.~Boland, ``Efficient encoding and decoding of binaural sound with resonance audio,'' in \emph{Audio Engineering Society Conference: 2019 AES International Conference on Immersive and Interactive Audio}.\hskip 1em plus 0.5em minus 0.4em\relax Audio Engineering Society, 2019.

\bibitem{google2023bert}
Google, ``Bert experts (wikibooks),'' \url{https://tfhub.dev/google/experts/bert/wiki_books/2}, 2023, accessed: January 10, 2025.

\bibitem{perez2018film}
E.~Perez, F.~Strub, H.~De~Vries, V.~Dumoulin, and A.~Courville, ``Film: Visual reasoning with a general conditioning layer,'' in \emph{Proceedings of the AAAI conference on artificial intelligence}, vol.~32, no.~1, 2018.

\bibitem{he2016deep}
K.~He, X.~Zhang, S.~Ren, and J.~Sun, ``Deep residual learning for image recognition,'' in \emph{Proceedings of the IEEE conference on computer vision and pattern recognition}, 2016, pp. 770--778.

\bibitem{allen1979image}
J.~B. Allen and D.~A. Berkley, ``Image method for efficiently simulating small-room acoustics,'' \emph{The Journal of the Acoustical Society of America}, vol.~65, no.~4, pp. 943--950, 1979.

\bibitem{muller1959note}
M.~E. Muller, ``A note on a method for generating points uniformly on n-dimensional spheres,'' \emph{Communications of the ACM}, vol.~2, no.~4, pp. 19--20, 1959.

\bibitem{de2015modeling}
E.~De~Sena, N.~Antonello, M.~Moonen, and T.~Van~Waterschoot, ``On the modeling of rectangular geometries in room acoustic simulations,'' \emph{IEEE/ACM Transactions on Audio, Speech, and Language Processing}, vol.~23, no.~4, pp. 774--786, 2015.

\bibitem{kahn2020libri}
J.~Kahn, M.~Riviere, W.~Zheng, E.~Kharitonov, Q.~Xu, P.-E. Mazar{\'e}, J.~Karadayi, V.~Liptchinsky, R.~Collobert, C.~Fuegen \emph{et~al.}, ``Libri-light: A benchmark for asr with limited or no supervision,'' in \emph{ICASSP 2020-2020 IEEE International Conference on Acoustics, Speech and Signal Processing (ICASSP)}.\hskip 1em plus 0.5em minus 0.4em\relax IEEE, 2020, pp. 7669--7673.

\bibitem{drossos2020clotho}
K.~Drossos, S.~Lipping, and T.~Virtanen, ``Clotho: An audio captioning dataset,'' in \emph{ICASSP 2020-2020 IEEE International Conference on Acoustics, Speech and Signal Processing (ICASSP)}.\hskip 1em plus 0.5em minus 0.4em\relax IEEE, 2020, pp. 736--740.

\bibitem{fonseca2021fsd50k}
E.~Fonseca, X.~Favory, J.~Pons, F.~Font, and X.~Serra, ``Fsd50k: an open dataset of human-labeled sound events,'' \emph{IEEE/ACM Transactions on Audio, Speech, and Language Processing}, vol.~30, pp. 829--852, 2021.

\bibitem{freesound}
``Freesound - {Collaborative} database of creative-commons licensed sound effects,'' \url{https://freesound.org/}, 2025, accessed: January 10, 2025.

\bibitem{zen2019libritts}
H.~Zen, V.~Dang, R.~Clark, Y.~Zhang, R.~J. Weiss, Y.~Jia, Z.~Chen, and Y.~Wu, ``Libritts: A corpus derived from librispeech for text-to-speech,'' \emph{arXiv preprint arXiv:1904.02882}, 2019.

\bibitem{gemmeke2017audio}
J.~F. Gemmeke, D.~P. Ellis, D.~Freedman, A.~Jansen, W.~Lawrence, R.~C. Moore, M.~Plakal, and M.~Ritter, ``Audio set: An ontology and human-labeled dataset for audio events,'' in \emph{2017 IEEE international conference on acoustics, speech and signal processing (ICASSP)}.\hskip 1em plus 0.5em minus 0.4em\relax IEEE, 2017, pp. 776--780.

\bibitem{shimada2024starss23}
K.~Shimada, A.~Politis, P.~Sudarsanam, D.~A. Krause, K.~Uchida, S.~Adavanne, A.~Hakala, Y.~Koyama, N.~Takahashi, S.~Takahashi \emph{et~al.}, ``Starss23: An audio-visual dataset of spatial recordings of real scenes with spatiotemporal annotations of sound events,'' \emph{Advances in Neural Information Processing Systems}, vol.~36, 2024.

\bibitem{lluis2023direction}
F.~Lluis, N.~Meyer-Kahlen, V.~Chatziioannou, and A.~Hofmann, ``Direction specific ambisonics source separation with end-to-end deep learning,'' \emph{Acta Acustica}, vol.~7, p.~29, 2023.

\end{thebibliography}
